\documentstyle[12pt,psfig]{article}
\begin{document}
\title{{\bf Generalised definitions of certain functions and their uses}}
\author{Debasis Biswas$^1$, S. Biswas$^{1,3}$\\
and\\
 Asoke P. Chattopadhyay$^2$ \\
$^1$Department of Physics, $^2$Department of Chemistry\\
University of Kalyani,\\
Kalyani  741235,\\
India\\
$^1$dbiswas-chak@rediffmail.com, $^2$asoke@klyuniv.ernet.in\\
$^3$sbiswas@klyuniv.ernet.in}
\maketitle
\begin{abstract}
Generalised definitions of exponential, trigonometric sine and cosine and
hyperbolic sine and cosine functions are given. In the lowest order, these
functions correspond to ordinary exponential, trigonometric sine etc. Some
of the properties of the generalised functions are discussed. Importance of
these functions and their possible applications are also considered.
\end{abstract}
\vskip 1.0cm
PACS No.: 02.30.Hq, 02.30.Gp.  MSC-class: 34A25; 34A34; 33B10. 
\newpage
{\section{Introduction}}
Ordinary differential equations (ODEs) are ubiquitous in various branches of 
physical and biological sciences, and are essential to our understanding 
of topics as diverse as stability theory, analysis of electrical signals and 
networks, chaos and nonlinear dynamics, radiative processes, 
etc.\cite{Bateman,Murphy,Ince,Hartman,Fors,Abram,Grad,Zwill,CRC}. In this paper we try to generalize the definition of $sine,\,cosine$ and $hyperbolic$ functions as solutions of differential equation of the form $y_{xx}+f(x)y=0$. For $f(x)\propto x, x^{-1}, x^2, x^{-2} $ the equations can be brought to Airy type, Coulomb type, Harmonic oscillator type and Euler type equations having wide application in physical science problems. We show that many well known ODE solutions of physical science can also be defined in terms of these generalized functions which we call now as $g$ functions. 
\par
The organization of the paper is as follows. In section 2 we formulate the definition of $g$ functions related to generalised exponential, generalised cosine and sine and generalised cosine hyperbolic and sine hyperbolic functions. In section 3 we discuss some important ODEs of mathematical physics in terms of $g$ functions.  In section 4 we end up with a concluding discussion.
 {\section{Formulation of $g$ function}} 
We begin with formal definitions of exponential, trigonometric and hyerbolic
functions as follows. The exponential function may be defined as solution of
the differential equation\\
\begin{equation}
\frac{dy}{dx} = y 
\end{equation}
Similarly, trigonometric sine and cosine functions may be defined as solutions
to the following differential equation\\
\begin{equation}
\frac{{d^2}y}{dx^2} = -y
\end{equation}
while hyperbolic sine and cosine functions are defined as solutions of\\
\begin{equation}
\frac{{d^2}y}{dx^2} = y
\end{equation}
Let us consider solutions to the following set of differential equations\\
\begin{eqnarray}
\frac{dy}{dx} &=& {x^n}y\\
\frac{{d^2}y}{dx^2} &=& -{x^n}y\\
\frac{{d^2}y}{dx^2} &=& {x^n}y
\end{eqnarray}
{\bf Case I : Generalised exponential function}
\par
Following Forsyth\cite{Fors}, we express gerenal solutions of (4) as\\
\begin{equation}
y = 1 + SP + SPSP + SPSPSP +......
\end{equation}
where S denotes the operation of integration from 0 to x i.e. $\int_0^x dx$ 
and $P = x^n$. Then the solution of (4) becomes\\
\begin{equation}
y = 1 + \frac{x^{(n+1)}}{(n+1)} + \frac{x^{(2n+2)}}{(n+1)(2n+2)} + \frac{x^{(3n+3)}}{(n+1)(2n+2)(3n+3)} + ...
\end{equation}
The series on RHS of (8) is convergent for finite x and n. The only 
restriction on n is that $n \ne -1$. We can also 
solve equation (4) by separation of variables. The result is\\
\[
lny = \frac{x^{(n+1)}}{(n+1)}
\]
so that\\
\begin{equation}
y = e^{\frac{x^{(n+1)}}{n+1}}
\end{equation}
It is easily verified that the RHS of the two expressions, (8) and (9), are
identical for finite x and n. These can be used to define exponential
function of order n as\\
\begin{equation}
e_{n}(x) = 1 + \frac{x^{(n+1)}}{(n+1)} + \frac{x^{(2n+2)}}{(n+1)(2n+2)} + \frac{x^{(3n+3)}}{(n+1)(2n+2)(3n+3)} + ...
\end{equation}
Note that the ordinary exponential function is zeroth order of this generalised
exponential function $e_{n}(x)$. Indeed, with n = 0, the above expression
becomes\\
\[
e_{0}(x) = e^x = 1 + \frac{x}{1!} + \frac{x^2}{2!} + \frac{x^3}{3!} + ....
\]
However, the generalised exponential allows expressions for other values of n.
Behaviour of $e_{n}(x)$ for some values of n are shown in Figure 1.\\
{\bf Case II : Generalised cosine and sine function}
\par
Let us solve equation (5) with the same technique. We get two independent
solutions as\\
\\
$y_1=1 + S^{2}P + S^{2}PS^{2}P + S^{2}PS^{2}PS^{2}P$ + ....\\
and\\
$y_2=x + S^{2}Px + S^{2}PS^{2}Px + S^{2}PS^{2}PS^{2}Px$ + ....\\
\\
Here S = $\int_0^x dx$ as before and P = $-x^n$. Then these solutions define
generalised cosine and sine functions for us as\\
\begin{eqnarray}
g^{n}_{c}(x) = 1 - {\frac{x^{(n+2)}}{(n+1)(n+2)}} + {\frac{x^{(2n+4)}}{(n+1)(n+2)(2n+3)(2n+4)}} -+...\\
g^{n}_{s}(x) = x - {\frac{x^{(n+3)}}{(n+2)(n+3)}} + {\frac{x^{(2n+5)}}{(n+2)(n+3)(2n+4)(2n+5)}} -+...
\end{eqnarray}
\\
We note that the functions $g^{n}_{c}(x)$ and $g^{n}_{s}(x)$, as given above, 
are not defined for negative integral (and for some negative rational) values of n. For n = 0, these 
functions are identical with ordinary trigonometric cosine and sine functions 
respectively. But expressions for other values of n are allowed. Behaviour of 
these functions for some values of n are shown in Figures 2 and 3 respectively.\\
\par
We now extend the above definitions for $g^{n}_{c}(x)$ and $g^{n}_{s}(x)$ for 
negative values of n as follows. $g^{-m}_c(x) = x g^{m-4}_s({\frac{1}{x}})$ leads to\\
\begin{eqnarray}
g^{-n}_{c}(x) = 1 - {\frac{x^{(2-n)}}{(n-1)(n-2)}} + {\frac{x^{(4-2n)}}{(n-1)(n-2)(2n-3)(2n-4)}} -+...\\
g^{-n}_{s}(x) = x - {\frac{x^{(3-n)}}{(n-2)(n-3)}} + {\frac{x^{(5-2n)}}{(n-2)(n-3)(2n-4)(2n-5)}} -+...
\end{eqnarray}
\\
Note that these expressions of $g^{-n}$ are also trivially obtained by 
substituting -n for n in corresponding expressions for $g^{n}$. It must be
noted that  $g^n_{s,c}$ are not defined for all negative values of n,
while $g^{-n}_{c}(x)$ is not defined for n  =1 and 2, $g^{-n}_{s}(x)$ is not
defined for n = 2 and 3. These functions are plotted for some negative n
values in Figures 4 and 5.\\
{\bf Case III: Generalised hyperbolic function}
\par
For $P = x^n$, all the signs in the two series above becomes positive, and
we get as solutions of equation (6),\\
\begin{eqnarray}
g^{n}_{hc}(x) = 1 + {\frac{x^{(n+2)}}{(n+1)(n+2)}} + {\frac{x^{(2n+4)}}{(n+1)(n+2)(2n+3)(2n+4)}} +...\\
g^{n}_{hs}(x) = x + {\frac{x^{(n+3)}}{(n+2)(n+3)}} + {\frac{x^{(2n+5)}}{(n+2)(n+3)(2n+4)(2n+5)}} +...
\end{eqnarray}
\\
where we have identified these solutions as generalised hyperbolic cosine and
sine functions respectively. Note that these functions are also not defined for
$n \le 0$ integers. It is trivial to show that $g^{0}_{hc}(x) = cosh(x)$ and
$g^{0}_{hs}(x) = sinh(x)$. $g^{n}_{hc}(x)$ is plotted for some values of n in
Figure 6. The behaviour of $g^{n}_{hs}(x)$, for the same n values, is very similar
and is therefore not shown separately.\\
\par
We extend these definitions for generalised hyperbolic functions to negative n
values in the same manner as was done for generalised trigonometric functions.
In other words\\
\begin{eqnarray}
g^{-n}_{hc}(x) = 1 + {\frac{x^{(2-n)}}{(n-1)(n-2)}} + {\frac{x^{(4-2n)}}{(n-1)(n-2)(2n-3)(2n-4)}} +...\\
g^{-n}_{hs}(x) = x + {\frac{x^{(3-n)}}{(n-2)(n-3)}} + {\frac{x^{(5-2n)}}{(n-2)(n-3)(2n-4)(2n-5)}} +...
\end{eqnarray}
\\
define the generalised hyperbolic functions for negative values of n. As 
before, these expressions can also be obtained from the corresponding
expressiosn for positive n values by substituting n with -n. The same 
restrictions as in the case of $g^{-n}_{c}$ and $g^{-n}_{s}$ apply for these
functions as well i.e. $n \ne 1, 2$ for $g^{-n}_{hc}(x)$ and $n \ne 2, 3$ 
for $g^{-n}_{hs}(x)$. Behaviour of these functions for some negative n values 
are shown in Figures 7 and 8. 
\par
It is clear that the above generalised definitions hold for positive and 
negative real values of n (except those mentioned above). 
Solutions to the second order ODEs with $-x^{n}y$ on RHS are oscillatory.
The oscillatory nature of these generalised functions is tested as follows. 
Leighton's oscillatory theorem\cite{Agar} asserts that a differential equation\\
\[
(p(x)y') + q(x)y = 0
\]
\\
is oscillatory in $(0,{\infty})$ provided $\int_0^{\infty}{\frac{1}{p(x)}}dx \rightarrow {\infty}$ 
and 
$\int_0^{\infty} q(x)dx \rightarrow {\infty}$. As both these conditions are satisfied 
in this case, the ODEs in (5) and (6) and hence their solutions are 
oscillatory.\\
\par
It can be proved that $g^{n}_{c}(x)$ and $g^{n}_{s}(x)$, as also
$g^{n}_{hc}(x)$ and $g^{n}_{hs}(x)$ are mutually orthogonal in 
$(-{\infty},{\infty})$. This shall be dealt with in a later communication.
We now propose that the functions $e_{n}(x)$, $g^{n}_{c}(x)$, 
$g^{n}_{s}(x)$, $g^{n}_{hc}(x)$ and $g^{n}_{hs}(x)$ be considered as 
generalised exponential, generalised cosine and sine, and generalised
hyperbolic cosine and sine functions respectively. From
plots of these functions in Figures 1 to 5, identity of these functions with
ordinary exponential, trigonometric cosine and sine, and hyperbolic cosine
and sine functions for n = 0, and departure from the latter functions for higher
values of n are evident.
{\section {Applications}}
{\bf (I)} We next show the utility of these functions. Consider any second order ODE of 
the form
\begin{equation}
\frac{{d^2}u}{dx^2} + 2R{\frac{du}{dx}} + (R^{2} + {\frac{dR}{dx}} + x^{n})u = 0
\end{equation}

Substituting $u = ye^{-{\int Rdx}}$, R=R(x), the above ODE becomes identical 
with equation (5), whose solutions (except for negative integral n) can be 
written in terms of $g^{n}_{c}(x)$ and $g^{n}_{s}(x)$. Therefore (except for some 
$n \le o$ integers), solutions of (19) can be expressed in terms of the latter 
functions.\\

{\bf (II)} These oscillatory functions can also serve as solutions to the following type 
of ODEs\\
\[
\frac{{d^2}y}{dx^2} + f(x)y = 0
\]
\\
where $f(x) = Ax^m + Bx^n +$ .... where A, B, .. are constants and m, n, ... 
may be relatively coprime.\\

{\bf (III)} Those familiar with second order ODEs can see the connection between equations
(5) and (6) and well known ODEs such as Bessel's, Airy's, Mathieu's etc. 
We discuss briefly on these connections, details will be placed in future works, with applications of $g^{n}_{c}(x)$ and 
$g^{n}_{s}(x)$ wherever possible. Here, we shall show how these functions 
can be used as solutions of Bessel's equation. Bessel's ODE is written as\\
\[
{x^2}{\frac{{d^2}y}{dx^2}} + x{\frac{dy}{dx}} + (x^{2} - \nu^{2}) y = 0
\]
\\
Putting $x = 2{\nu}r^{\frac{1}{2{\nu}}}$ and $y = ur^{-{\frac{1}{2}}}$, 
the above ODE becomes\\
\[
\frac{{d^2}u}{dr^2} + r^{\frac{1}{{\nu}-2}}u = 0
\]
\\
the solution of which can be written as\\
\[
u(r) = A g^{\frac{1}{{\nu}-2}}_{c}(r) + B g^{\frac{1}{{\nu}-2}}_{s}(r)
\]
\\
Hence, the general solution of Bessel's ODE can be expressed as\\
\[
y(x) = ({\frac{2{\nu}}{x}})^{\nu}[A g^{\frac{1}{{\nu}-2}}_{c}{({\frac{x}{2{\nu}}})^{2{\nu}}} + B g^{\frac{1}{{\nu}-2}}_{s}{({\frac{x}{2{\nu}}})^{2{\nu}}}]
\]
\\
{\bf (IV)} Let us discuss the solution of Ricatti equation in terms of $g$ function. The Ricatti equation is
\begin{equation}
\frac{dy}{dx}+y^2+x^m=0
\end{equation}
With $\frac{u^\prime}{u}=y$, the above equation can be cast into the form
\begin{equation}
\frac{u^{\prime\prime}}{u}+x^m=0
\end{equation}
which evidently has two solutions in terms of $g_s^n$ and $g_c^n$.\\

{\bf  (V) }
The confluent hypergeometric equation is 
\begin{equation}
x\frac{d^2y}{dx^2}+(k-x)\frac{dy}{dx}-ay=0
\end{equation}
If we substitute $x=-t^2/4$, $y=t^{-n}z$ with $k=n+1$, the confluent hypergeometric equation is reduced to Bessel differential equation
\begin{equation}
t^2\frac{d^2z}{dt^2}+t\frac{dz}{dt}+(t^2-n^2)z=0
\end{equation}
and can be solved in terms of $g_s^n$ and $g_c^n$ as shown earlier.
\par
Moreover Whittaker equation and Weber-Hermite equation can also be solved in terms of $g_s^n$ and $g_c^n$. 
The above examples merely shows the equivalence of the method adopted in this
work with one that is much more familiar in physical science. 
\par
We now present an example where the present method is shown to be superior to existing ones. In the case of an
ODE such as\\
\[
\frac{d^{2}y}{dx^{2}} - x^{-4}y = 0
\]
\\
 no regular solution can be obtained by the usual, Frobenius method. Only
normal integrals exist in such cases \cite{Fors}. But the present method 
immediately provides regular solutions in a straightforward manner. This is
also apparent in cases where logarithmic terms appear in the Frobenius method
which are troublesome to deal with. Such a possibility does not arise in the
present method. Also, for complicated nonlinear ODEs for example, often one
solution appears in published tables of integrals; the present method yields
all possible solutions. We just cite one example.\\ 
\par
Consider the equation
\begin{equation}
yy^{\prime\prime\prime}+3y^\prime y^{\prime\prime}+x^myy^\prime=0
\end{equation}
With substitution $w=yy^\prime$, the above equation reduces to the form
\begin{equation}
w^{\prime\prime}+x^mw=0
\end{equation}
Obviously the solutions are $w_1=g_s^m(x)$ and $w_2=g_c^m(x)$.
{\section{Concluding discussion}}
Applicability of the generalised exponential and the generalised hyperbolic
functions defined above can also be established in a similar manner. It
becomes apparent that such functions can be applied to a fairly broad category
of physical problems. While the present work may be considered as a brief 
introduction to these functions, their detailed nature and 
properties will be taken up in our future works. We hope to 
establish the usefulness of these functions, and show how solutions of various 
the problems of mathematical physics can be expressed in much more compact
and uniform manner with their help.\\
\par
We wish to emphasise that the solutions to a class of differential equations
given in the present work is actually a subset of a larger class of functions.
Here, we have only considered first and second order ODEs of a particular
type. We wish to state that these are subsets of the following ODEs viz.\\
\begin{equation}
\frac{d^{n}y}{dx^{n}} = \pm x^{m}y
\end{equation}
\\
where n and m may both vary. 
We wish to express solutions to such ODEs as $g^{\pm,k}_{n,m}(x)$, where
k = 0, 1, ... n. For
example, $e_{n}(x)$ is now written as $g^{+,1}_{1,0}(x)$, $g^{n}_{c}(x)$ is 
written as $g^{-,1}_{2,n}(x)$, $g^{n}_{s}(x)$ is written as $g^{-,2}_{2,n}(x)$, 
$g^{n}_{hc}(x)$ is written as $g^{+,1}_{2,n}(x)$ and $g^{n}_{hs}(x)$ is
written as $g^{+,2}_{2,n}(x)$. While the notation used earlier was adequate
for denoting solutions to lower order ODEs as shown at the beginning of this
work, viz. equations (4) to (6), the need for a more general notation becomes 
immediately apparent as higher order ODEs are considered. For example, for\\
\[
\frac{d^{3}y}{dx^{3}} = x^{m}y
\]
\\
the solutions can be written as $g^{+,1}_{3,m}(x)$, $g^{+,2}_{3,m}(x)$ and
$g^{+,3}_{3,m}(x)$. Such functions will be considered in our future 
publications.

{\bf{Acknowledgements:}}
The authors wish to acknowledge the help received from Prof. P. Dutt and 
Prof. S. Madan of Department of Mathematics, I.I.T., Kanpur, India.

\newpage

\begin{figure}[t]
\psfig{figure=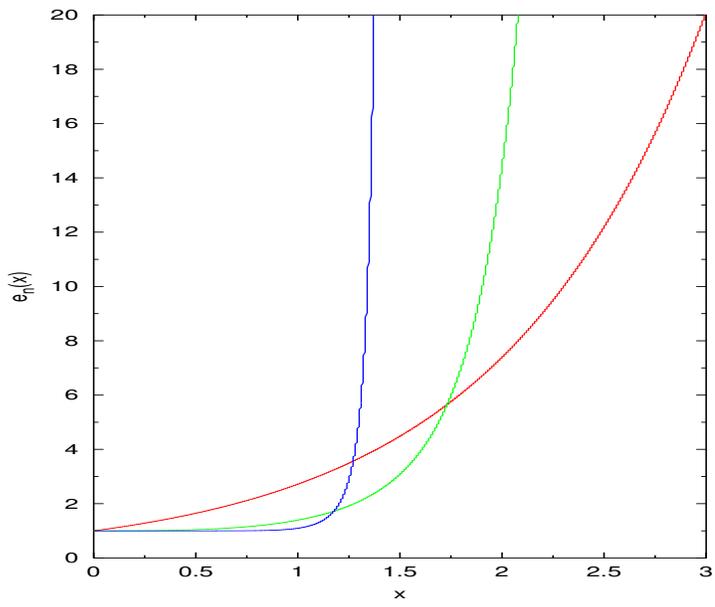,height=8cm,width=10cm}
\caption{$e_{n}(x)$ vs x for n = 0 (red), 2 (green) and 10 (blue).}
\end{figure}
\newpage
\begin{figure}[t]
\psfig{figure=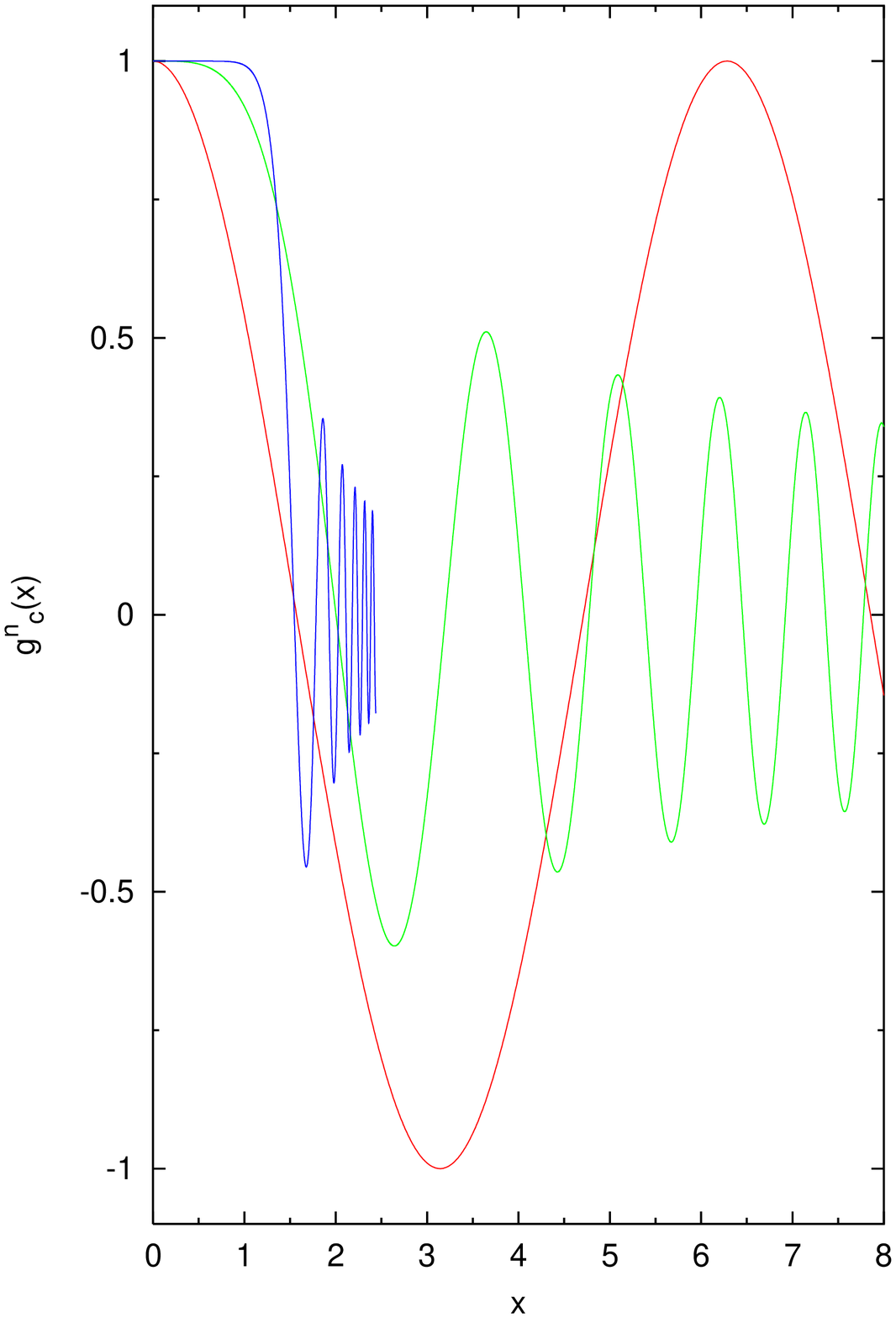,height=6in,width=5.5in}
\caption{$g^{n}_{c}(x)$ vs x for n = 0 (red), 2 (green) and 10 (blue). 
Note increased oscillation as n is increased. The abrupt termination 
for the blue curve is due to numerical instability.}
\end{figure}
\newpage
\begin{figure}[t]
\psfig{figure=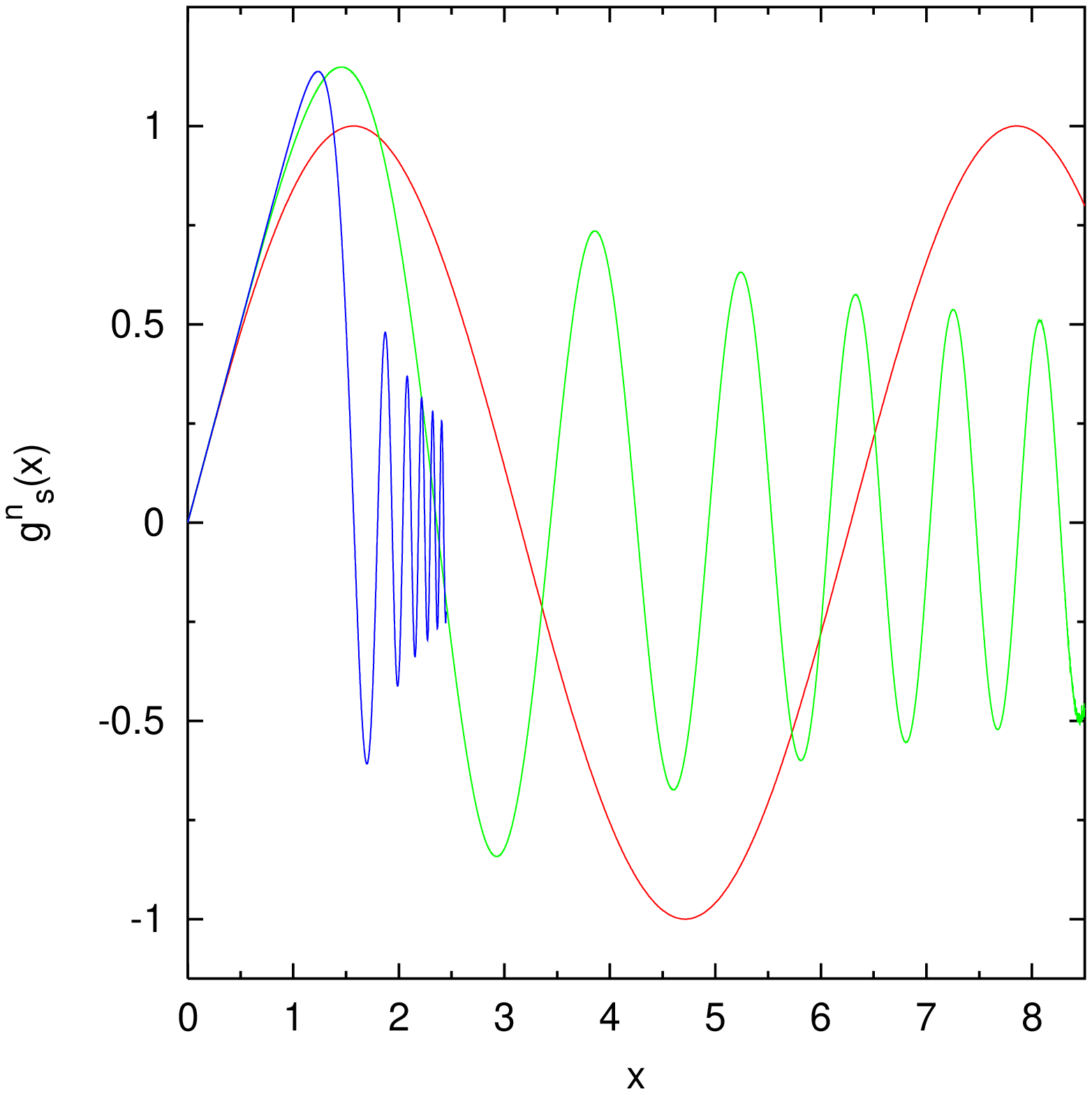,height=6in,width=5.5in}
\caption{$g^{n}_{s}(x)$ vs x for n = 0 (red), 2 (green) and 10 (blue). 
Note increased oscillation as n is increased. The abrupt termination 
for the blue curve is due to numerical instability.}
\end{figure}
\newpage
\begin{figure}[t]
\psfig{figure=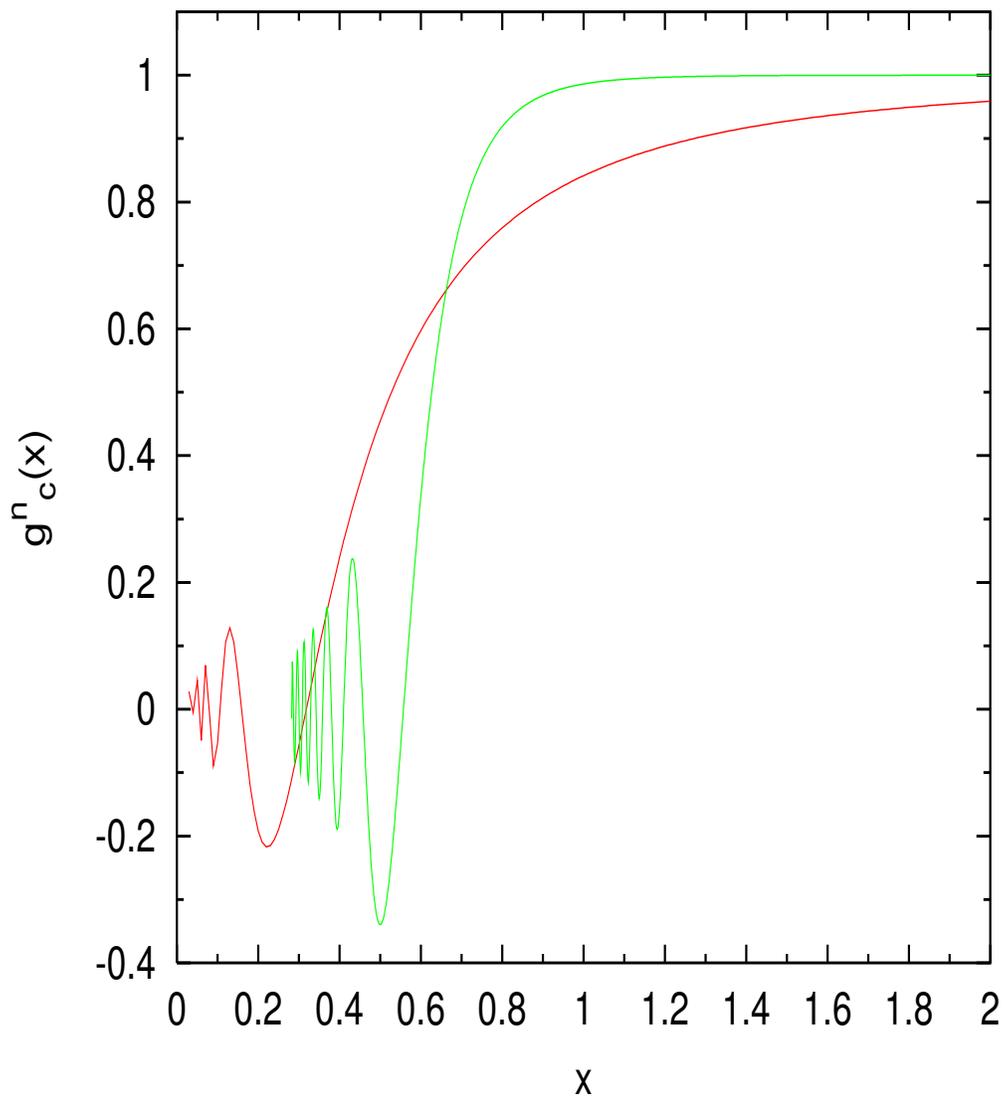,height=6in,width=5.5in}
\caption{$g^{n}_{c}(x)$ vs x for n = -4 (red) and -10 (green). 
Note oscillatory behaviour for small x and asymptotic behaviour for 
large x as expected from equation (13). Oscillations increases as magnitude 
of n is increased. The abrupt termination, more apparent for n = -10 is due to 
numerical instability.}
\end{figure}
\newpage
\begin{figure}[t]
\psfig{figure=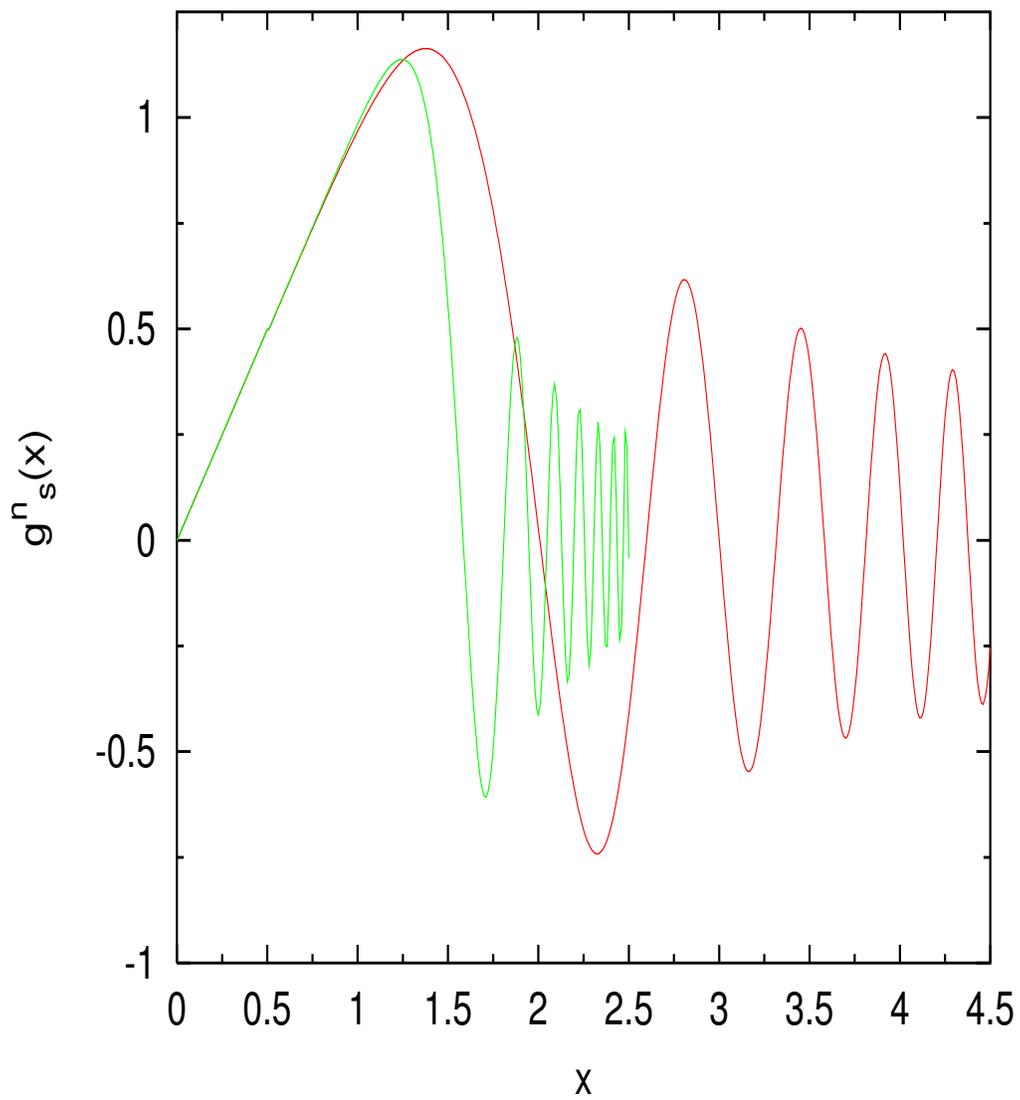,height=6in,width=5.5in}
\caption{$g^{n}_{s}(x)$ vs x for n = -4 (red) and -10 (green). Note that 
for small x, $g^{n}_{s}(x) \sim x$ as expected from equation (14). For larger 
x, the function oscillates, more for n = -10. The abrupt termination is due to
numerical instability.}
\end{figure}
\newpage
\begin{figure}[t]
\psfig{figure=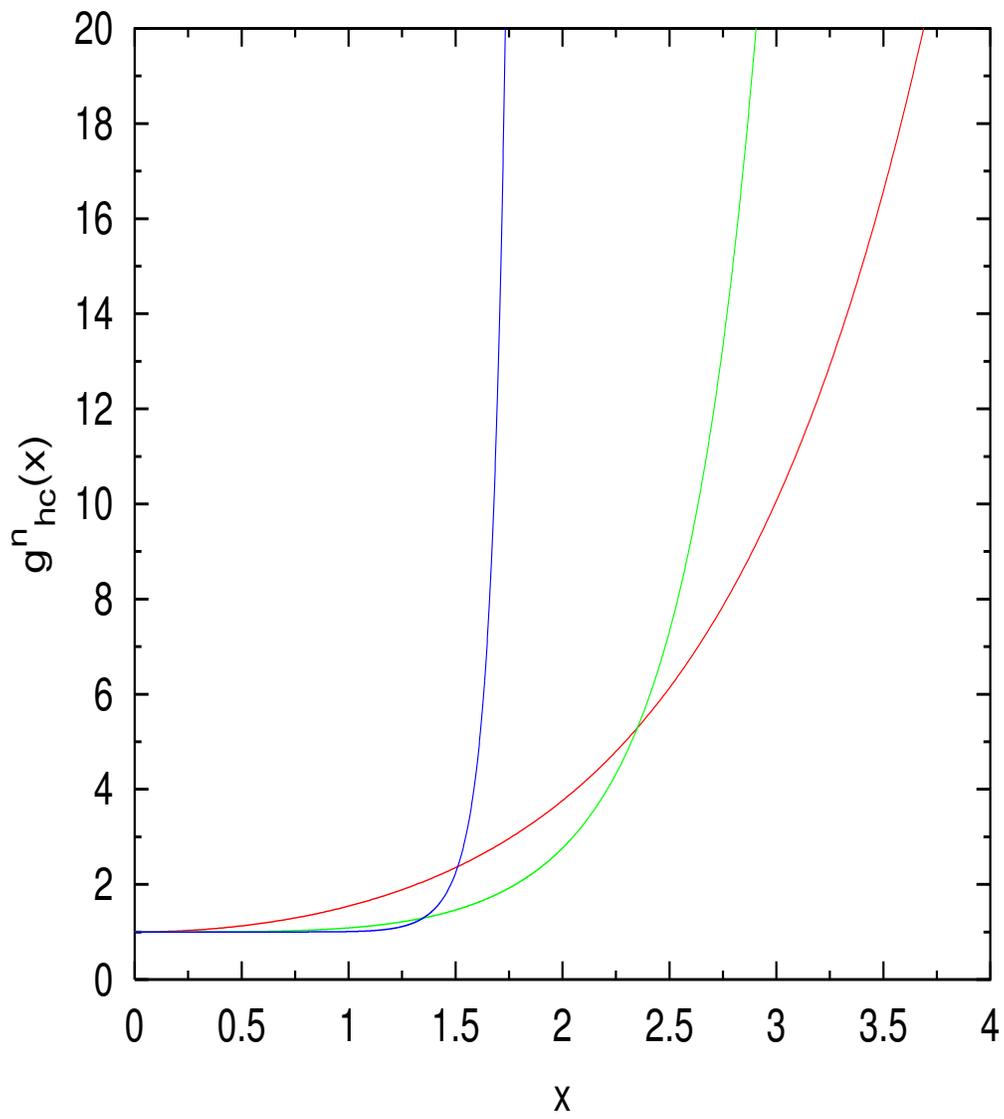,height=6in,width=5.5in}
\caption{$g^{n}_{hc}(x)$ vs x for n = 0 (red), 2 (green) and 10 (blue).}
\end{figure}
\newpage
\begin{figure}[t]
\psfig{figure=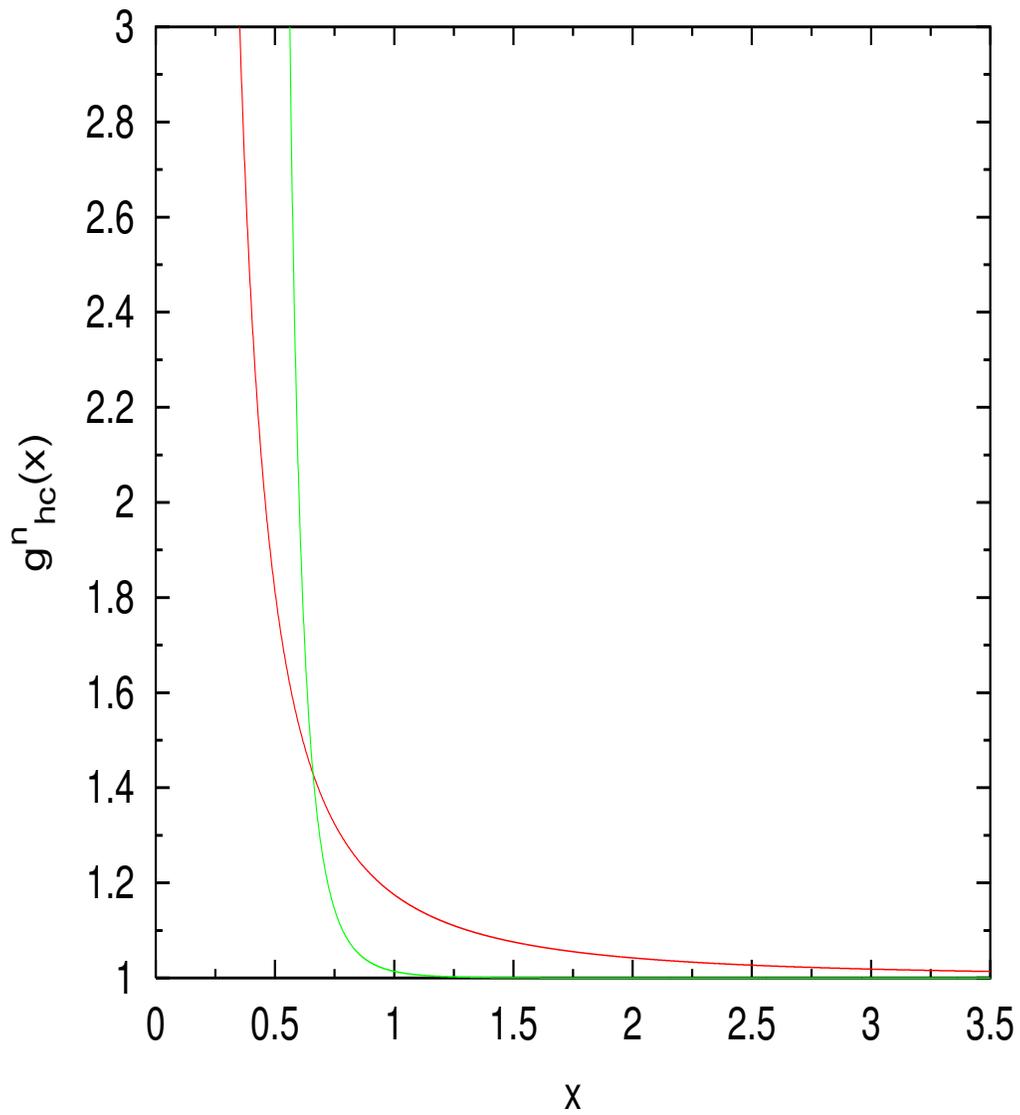,height=6in,width=5.5in}
\caption{$g^{n}_{hc}(x)$ vs x for n = -4 (red) and -10 (green). Note the
instability near x = 0 and smooth approach towards unity (first term in the
series) for large x, as expected.}
\end{figure}
\newpage
\begin{figure}[t]
\psfig{figure=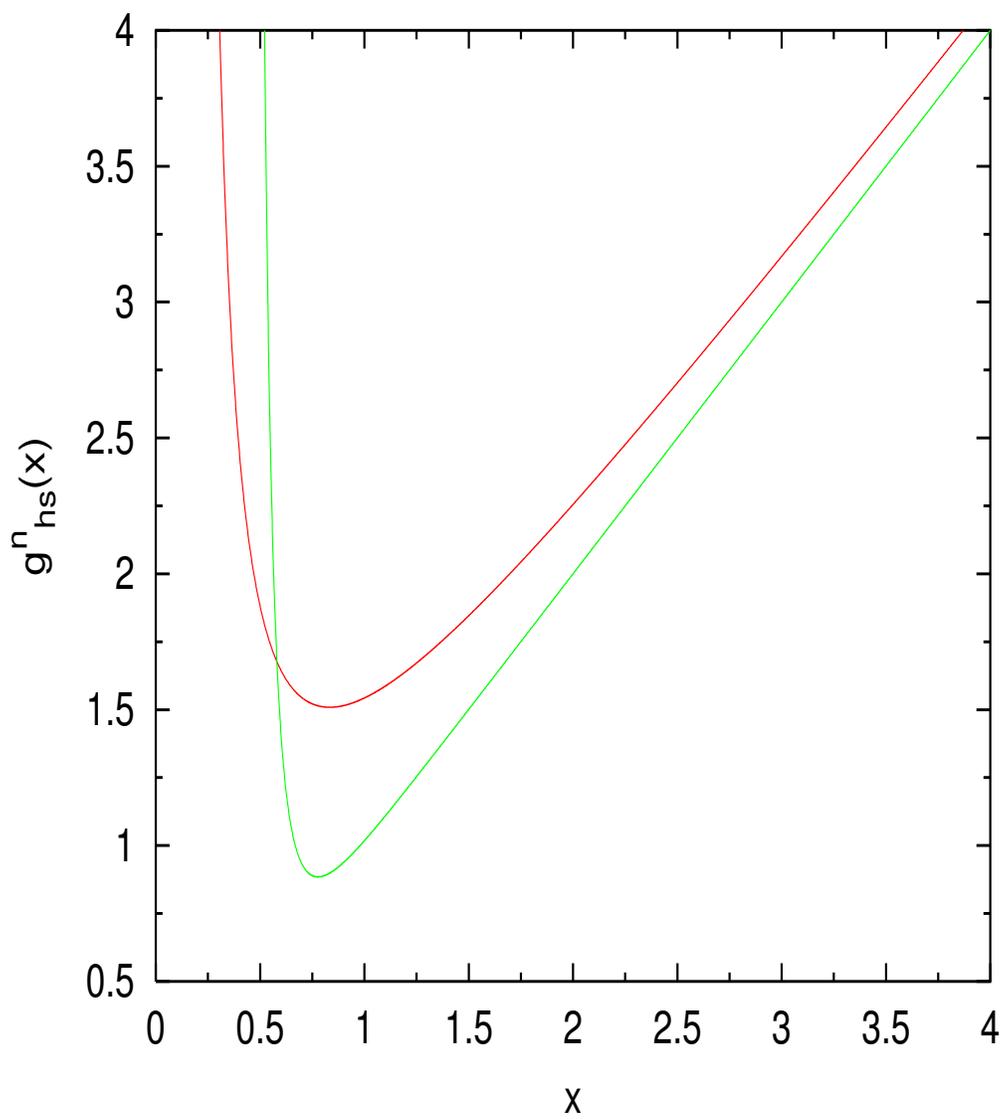,height=6in,width=5.5in}
\caption{$g^{n}_{hs}(x)$ vs x for n = -4 (red) and -10 (green). Note the
instability near x = 0. For large x, $g^{n}_{hs}(x) \sim x$ as expected (see
also Figure 5).}
\end{figure}
\end{document}